\documentclass[prl,amsmath,amssymb,twocolumn,superscriptaddress,showpacs]{revtex4}

\usepackage{amsmath,amssymb}
\usepackage[usenames]{color}
\usepackage{amssymb}
\usepackage{grffile}
\usepackage[pdftex]{graphicx}
\usepackage{amsmath, amstext, amssymb, amsfonts, amsxtra}
\usepackage{textcomp}
\usepackage{xspace}

\def \ell{{d}}

\newcommand{\bop}{\hat{b}} 
\newcommand{\nop}{\hat{n}} 
\newcommand{\bdop}{\hat{b}^{\dagger}}

\newcommand{\sutd}{Singapore University of Technology and Design, 20 Dover Drive, 138682 Singapore}
\newcommand{\hannover}{Institut f\"ur Theoretische Physik, Leibniz Universit\"at Hannover, Appelstr. 2, DE-30167 Hannover, Germany}
\newcommand{\majulab}{Merlion MajuLab, CNRS-UNS-NUS-NTU International Joint Research Unit, UMI 3654, Singapore}

\begin{document}

\title{Exploring Unconventional Hubbard Models with Doubly Modulated Lattice Gases}

\author{Sebastian Greschner} 
\affiliation{\hannover}
\author{Luis Santos}
\affiliation{\hannover}   
\author{Dario Poletti}
\affiliation{\sutd}
\affiliation{\majulab}

\begin{abstract}
Recent experiments show that periodic modulations of cold atoms in optical lattices may be used to engineer and explore interesting models.
We show that double modulation, combining lattice shaking and modulated interactions allows for the engineering of a much broader class of 
lattice models with correlated hopping, which we study for the particular case of one-dimensional systems. 
We show, in particular, that by using this double modulation it is possible to study Hubbard models with asymmetric hopping, 
which, contrary to the standard Hubbard model, present insulating phases with both 
parity and string order. Moreover, double modulation allows for the simulation of lattice models in unconventional parameter regimes, as 
we illustrate for the case of the spin-$1/2$ Fermi-Hubbard model with correlated hopping, a relevant model for cuprate superconductors.
\end{abstract}


\maketitle



\emph{Introduction.--} Ultracold gases in optical lattices have attracted a lot of attention as emulators of fundamental models of 
quantum many-body systems. Their unprecedented levels of controllability, tunability, and cleanness has permitted the realization 
of Hubbard models with cold atoms~\cite{Greiner2002,Joerdens2008,Schneider2008}, the creation of synthetic magnetic fields in neutral lattice gases~\cite{Aidelsburger2013,Miyake2013}, 
first steps towards the emulation of quantum magnetism~\cite{Trotzky2008,Simon2011,Nascimbene2012,Greif2013,Meinert2013,Yan2013,dePaz2013}, 
and more~\cite{Lewenstein2007,Bloch2008}. Recent progress in measurement techniques 
has allowed for single-site resolved detection~\cite{Bakr2009,Sherson2010}, which has permitted a deeper insight 
into the properties of Mott insulators~\cite{Sherson2010,Bakr2010,Endres2011}, and the dynamical properties of cold lattice  gases~\cite{Cheneau2012}.

The possibility of tuning parameters in real time in cold lattice gases has aroused a particular interest. Fast periodic modulations provide a new tool for the engineering of relevant lattice models~\cite{Aidelsburger2013,Miyake2013,Eckardt2005,Lignier2007,Sias2008,Kierig2008,Zenesini2009,Struck2011,Chen2011,Ma2011,Struck2012,Chin2013}. In particular, a fast-enough modulation of the lattice position~(lattice shaking) results in the effective change of the tunneling rate~\cite{Eckardt2005} allowing, for example, driving the SF to MI transition \cite{Zenesini2009}, inducing photon-assisted hopping in tilted 
lattices~\cite{Sias2008}, simulating frustrated classical magnetism \cite{Struck2011}, generating gauge potentials \cite{Struck2012} and inducing effective ferromagnetic domains \cite{Chin2013}. Moreover, a fast modulation of the interparticle interactions results in an effective hopping that depends on the occupation differences at neighboring sites \cite{Gong2009,Abdullaev2010, Rapp2012,DiLiberto2013}, and may induce density-dependent gauge fields \cite{Greschner2013}. 

In this work we show that double modulation~(DM), i.e. the combination of lattice shaking and periodically modulated interactions, permits the selective control of different hopping processes, hence allowing for the engineering of a broad range of lattice models that cannot be realized with either lattice shaking or modulated interaction alone. 
DM permits, in particular, the generation of mirror-asymmetric tunneling, which may result in insulators with both parity and string order. Moreover, DM permits the exploration of parameter regimes which cannot be reached with a single modulation, as we illustrate for the relevant case of spin-$1/2$ lattice fermions with correlated hopping. 



\emph{Double modulation.--} We consider a lattice gas with periodic DM. Whereas lattice shaking results from the displacement of the 
lattice~\cite{Eckardt2005}, periodic interactions may be induced by modulating an externally applied magnetic field in the vicinity of a Feshbach resonance~(see Refs.~\cite{Gong2009,Rapp2012,DiLiberto2013} for details). 
We focus below on 1D lattices, although the engineering possibilities of DM may be extended to higher dimensions as well, and in 2D lattices elliptic shaking~\cite{Eckardt2010} may be employed to induce even richer lattice models. 
In the experimentally relevant scenario in which only the lowest Bloch band is relevant \cite{FootnoteHighBands}, Bose gases are described in the lattice reference frame by the time-dependent Bose-Hubbard model~(BHM):
\begin{equation} 
\hat{H}=-J\sum_{\langle i,j\rangle} \bdop_i \bop_j + \frac{U(t)} 2 \sum_j \nop_j(\nop_j-1)+ F(t)\sum_j j\nop_j, 
 \label{eq:Hamiltonian}       
\end{equation}
where $\bop_i$($\bdop_i$) annihilates~(creates) a boson at site $i$, $\nop_i=\bdop_i\bop_i$, $J$ is the tunneling parameter, $U(t)$ is the time-dependent interaction strength and $F(t)$ is a time-dependent tilting amplitude resulting from the lattice shaking. Both the interaction and the tilting term are periodically modulated, $U(t)=U_0+U_1 f_U(t)=U_0+U_1 f_U(t+T)$ and $F(t)=F_1 f_F(t)=F_1 f_F(t+T)$. Both $f_{U,F}(t)$ are unbiased, i.e. $\int_t^{t+T}f_{U,F}(\tau)d\tau=0$. If $\omega=2\pi/T\gg U_0/\hbar,\;J/\hbar$, Floquet analysis may be employed to integrate the modulations, obtaining  an effective time-independent Hamiltonian~(see e.g. Refs.~\cite{Eckardt2005, Rapp2012} or Ref.~\cite{DiLiberto2013} for an equivalent derivation for the Fermi-Hubbard model):  
\begin{equation} 
\hat{H}_{eff}=-J\sum_{\langle i,j\rangle} \bdop_i \mathcal{F}\left( i-j ,\nop_i-\nop_j \right)\bop_j  + \frac{U_0} 2 \sum_j \nop_j(\nop_j-1)
\label{eq:effHamiltonian}       
\end{equation}
where $\mathcal{F}=\frac 1 T \int_0^T e^{{\rm i}/\hbar \int_0^t \left[F_1 f_F(\tau)(i-j)+U_1 f_U(\tau)(\nop_i-\nop_j)  \right] d\tau} dt$. 
We illustrate below the engineering possibilities allowed by DM for the particular case of $f_U(t)=f_F(t)=\cos(\omega t)$. Note however that different frequencies and/or functional forms for the two modulations may allow for an even more versatile engineering. The above mentioned choice implies $\mathcal{F}=\mathcal{J}_0\left[ \frac{F_1}{\hbar\omega}(i-j) + \frac{U_1}{\hbar\omega} \left( \nop_i-\nop_j \right) \right]$ 
in Eq.~\eqref{eq:effHamiltonian}, where $\mathcal{J}_n$ is the $n$-th order Bessel function~\cite{footnote-1}. 


\emph{Hopping channels.--} We denote as $J_{(n_i,n_j)\leftrightarrow (n_i+1,n_j-1)}$ the hopping rate from site $j$ with $n_j$ particles before the hop to site $i$ with initially $n_i$ particles. In 1D lattices a hop to the left is characterized by the rate
$$
\frac{J_{(n_j,n_{j+1})\leftrightarrow (n_j+1,n_{j+1}-1)}}{J}\! =\! {\cal J}_0\left ( 
\frac{U_1}{\hbar\omega}(n_j\! -\!n_{j+1}\! +\! 1)\! -\! \frac{F_1}{\hbar\omega}
\right ),
$$
whereas a hop to the right is given by
$$
\frac{J_{(n_j,n_{j+1})\leftrightarrow (n_j-1,n_{j+1}+1)}}{J}\! =\! {\cal J}_0\left ( 
\frac{U_1}{\hbar\omega}(n_{j+1}\! -\!n_{j}\! +\! 1)\! +\! \frac{F_1}{\hbar\omega}
\right ),
$$
Note that for sole lattice shaking ($U_1=0$) hops are mirror symmetric since ${\cal J}_0$ is even. The same is true for solely modulating interactions~($F_1=0$). DM allows for breaking mirror symmetry. Especially relevant at low fillings are $J_{(0,1)\leftrightarrow (1,0)}=J_{(1,2)\leftrightarrow (2,1)}=J{\cal J}_0\left (\frac{F_1}{\hbar\omega}\right )$, $J_{(1,1)\leftrightarrow (2,0)}=J{\cal J}_0\left (\frac{U_1-F_1}{\hbar\omega}\right )$, and $J_{(1,1)\leftrightarrow (0,2)}=J{\cal J}_0\left (\frac{U_1+F_1}{\hbar\omega}\right )$. One may observe two important features: (i) contrary to the standard BHM~($F_1=U_1=0$), in general  $J_{(1,1)\leftrightarrow (2,0)}\neq J_{(1,1)\leftrightarrow (0,2)}$; and (ii) $J_{(1,1)\leftrightarrow (2,0)}/J_{(0,1)\leftrightarrow (1,0)}$ and/or $J_{(1,1)\leftrightarrow (0,2)}/J_{(0,1)\leftrightarrow (1,0)}$ may be larger than one, an impossibility for $U_1=0$ and/or $F_1=0$. These peculiar features have crucial consequences. As shown below,  (i) DM may result in insulators with both parity and string order, whereas (ii) it also allows for the study of a much richer phase diagram for lattice gases, compared to the case of either shaking or interaction modulation.



\emph{Insulators with finite parity and string order.--} In the standard BHM, the MI with unit occupation 
is characterized by doublon-holon pairs in a sea of singly-occupied sites. The pairs result from the 
$J_{(1,1)\leftrightarrow (2,0)}=J_{(1,1)\leftrightarrow (0,2)}$ hops. The 1D MI at unit filling presents nonlocal parity order 
${\cal O}^2_P\equiv\lim_{|i-j|\rightarrow\infty} \langle (-1)^{\sum_{i<l<j}\delta \hat n_l} \rangle >0$, with $\delta \hat n_j=1-\hat n_j$~\cite{Berg2008}, 
due to the appearance of doublon and holon defects in pairs, as recently revealed experimentally~\cite{Endres2011}. 
Another important nonlocal order in 1D is string order,  ${\cal O}^2_S\equiv\lim_{|i-j|\rightarrow\infty} -\langle\delta\hat n_i (-1)^{\sum_{i<l<j}\delta\hat n_l} \delta\hat n_j\rangle$~\cite{DallaTorre2006}. A non-vanishing ${\cal O}^2_S$ characterizes the Haldane insulator~(HI), predicted in polar lattice gases~\cite{DallaTorre2006} and bosons in frustrated lattices~\cite{Greschner2013b}. In the HI 
 the position of the defects and the separation between them is not fixed, but starting with a doublon the next defect along the chain is a holon, the next a doublon, and so on. This 
diluted ``antiferromagnetic'' order is characterized by ${\cal O}_S^2>0$. However, in a HI defects are not paired, and hence ${\cal O}^2_P=0$. Conversely, in the standard BHM 
the MI presents ${\cal O}_S^2=0$ due to the equal probability of having $20$ and $02$ pairs.

\begin{figure}[ht]
\centering
\includegraphics[width=\linewidth]{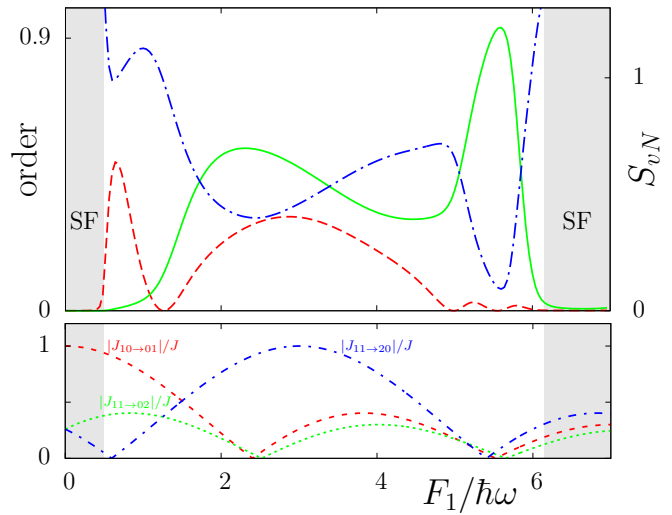} 
\caption{(Color online) (top) Infinite DMRG~\cite{Schollwock2011} results for the effective model~\eqref{eq:effHamiltonian} (with $4$ bosons per site, keeping up to $M=200$ states) for ${\cal O}_P^2$~(solid), $4 {\cal O}_S^2$~(dashed), and $S_{vN}$~(dot-dashed) as a function of $F_1/\hbar\omega$ for $U_0/J=1.2$, unit filling, and $U_1/\hbar\omega=3$. In the shaded SF-regions we find a logarithmic divergence of $S_{vN}$ with the number of kept matrix-states $M$. (bottom) Relevant hopping rates.}
\label{fig:1}
\end{figure}


DM allows for the selective reduction or cancellation of $J_{(1,1)\leftrightarrow (2,0)}$ (or alternatively  $J_{(1,1)\leftrightarrow (0,2)}$). As a result, 
inside the MI, pairs may be produced dominantly, or even solely, following a $02$ (or $20$) order. The system may then present both ${\cal O}^2_P>0$, due to the 
pairwise creation of defects, and ${\cal O}^2_S>0$, due to the dominantly ``antiferromagnetic'' order of the defects. This is best understood from the case in which 
all hoppings vanish except $J_{(1,1)\leftrightarrow (0,2)}$~\cite{footnote-2}. For unit filling, since the defect-free MI at $J=0$ just develops pairs $02$ at finite $J$, 
we map into an effective spin-$1/2$ system, with $|0,2\rangle_{j,j+1}\equiv |\uparrow\rangle_j$ and $|1,1\rangle_{j,j+1}\equiv |\downarrow\rangle_j$, 
obtaining an effective spin model:
\begin{eqnarray}
H_{S} &=& \sum_j J_{(1,1)\leftrightarrow (0,2)} \left(\hat S^+_j+ \hat S^-_j\right) + U_0 \left(\frac{1}{2} + \hat S^z_j\right) \nonumber \\ &+& \Delta \left(\frac{1}{2}
 + \hat S^z_j\right)\left(\frac{1}{2} + \hat S^z_{j+1}\right)
\label{eq:HI2}
\end{eqnarray}
where $\hat S_j^{\pm,z}$ are spin-$1/2$ operators, and we add a spin-spin interaction $\Delta\rightarrow\infty$ to project out neighboring up spins, which have no physical meaning. 
Interestingly, Hamiltonian~\eqref{eq:HI2} corresponds to an Ising model with transverse and longitudinal magnetic fields in the vicinity of its tricritical point between, ferro-, antiferro-, and paramagnetic phases~\cite{Ovchinnikov2003}. For a chain of length $L$ and periodic boundary conditions, the ground state of~\eqref{eq:HI2}
is a linear combination of states with $m$ spin ups, i.e. $m$ defect pairs, of the form $\sum_{m=0}^{L/2} c_m |m\rangle$, 
where $|m\rangle \propto (PS^+)^m |0\rangle$ with $P$ the projector excluding states with 
two neighboring up spins, $S^+=\sum_j S_j^+$, and $|0\rangle$ the state with all spins down (i.e. a defect-free MI).
String and parity order may then be expressed as statistical moments of the number of defect pairs: ${\cal O}^2_P \approx \left<(2m-1)^2\right>$ and ${\cal O}^2_S \approx \left<2m^2 (m-2)\right>$. Both orders may hence coexist. 
Note that with increasing number of pairs, i.e. decreasing ratio $U_0/J_{(1,1)\leftrightarrow (0,2)}$, ${\cal O}^2_S$ increases whereas ${\cal O}^2_P$ decreases, as 
observed in our simulations of Eq.~\eqref{eq:effHamiltonian}.


Figure~\ref{fig:1}(top) illustrates the rich physics that results from the selective control of the different hopping rates allowed by DM. 
For a fixed $U_0/J=1.2$ and $U_1/\hbar\omega=3$, we analyze, by means of density matrix renormalization group~(DMRG)~\cite{Schollwock2011} calculations, the 
different order parameters for a varying $F_1/\hbar\omega$. The relevant tunneling rates are
 depicted in Fig.~\ref{fig:1}~(bottom). One observes 
both that the hopping is generally asymmetric, $J_{(1,1)\leftrightarrow (2,0)}\neq J_{(1,1)\leftrightarrow (0,2)}$, and the existence of parameter regimes at which $J_{(0,1)\leftrightarrow (1,0)}<J_{(1,1)\leftrightarrow (2,0)}$ and/or $J_{(1,1)\leftrightarrow (0,2)}$. A first consequence of the variation of the hopping rates with $F_1$ is clearly 
the appearance of SF to insulator transitions 
(at $F_1/\hbar\omega\simeq 0.5$ and $6.2$)
. In the SF regime the excitation gap vanishes, the entanglement entropy $S_{vN}$ ~\cite{Eisert2010} shows a logarithmic divergence with the system-size~(not shown here), and correlations decay algebraically as expected for a Luttinger liquid. Within the insulator regions the entanglement entropy is finite and correlations decay exponentially. As 
expected from the discussion above, insulator phases with ${\cal O}_P^2>0$ and ${\cal O}_S^2>0$ occur due to the hopping asymmetry. Note that ${\cal O}_S^2$
increases when the hopping asymmetry grows, disappearing at those $F_1$ values at which $J_{(1,1)\leftrightarrow (2,0)}= J_{(1,1)\leftrightarrow (0,2)}$ (at $F_1/\hbar\omega\simeq 1.2, 5.2 \text{ and } 5.5$).
When $J_{(1,1)\leftrightarrow (0,2)}, J_{(1,0)\leftrightarrow (0,1)}\ll J_{(1,1)\leftrightarrow (2,0)}$, we recover the extreme case discussed above, and ${\cal O}_P^2$ and ${\cal O}_S^2$ 
are large (e.g. at $F_1/\hbar\omega\approx 2.4$). Interestingly, within the insulator regions, when the hopping is asymmetric and $J_{(1,1)\leftrightarrow (2,0)}, J_{(1,1)\leftrightarrow (0,2)}\ll J_{(1,0)\leftrightarrow (0,1)}$, 
we obtain ${\cal O}_P^2\ll {\cal O}_S^2$. This parameter region is characterized by, to a large extent, broken defect pairs but still ``antiferromagnetic'' defect order, i.e. in this region the insulator rather behaves as a HI. We find however no gapless region 
that would mark a transition between a HI~(with low ${\cal O}_P^2$) and a MI~(with finite ${\cal O}_S^2$)~\cite{Berg2008}.

\begin{figure}[t]
\centering
\includegraphics[width=\linewidth]{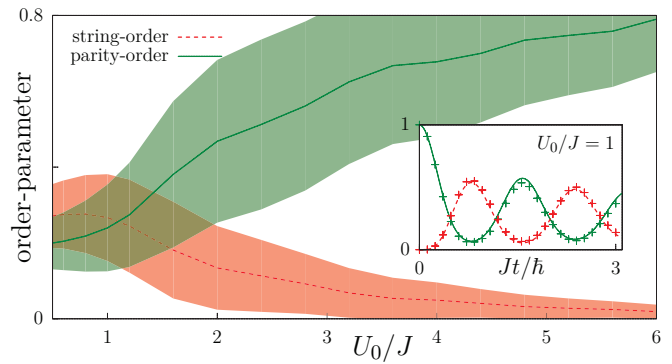}
\caption{(Color online) Dynamics of ${\cal O}_P^2$~(solid) and ${\cal O}_S^2$~(dashed) after a sudden onset of DM for $F_1/\hbar\omega=2.4$, $U_1/\hbar\omega=3$ for a system initially prepared as deep Mott-insulator. The inset shows the dynamics for $U_0/J=1$. The curves are obtained using the effective model~\eqref{eq:effHamiltonian}, whereas the crosses indicate the results directly obtained from Eq.~\eqref{eq:Hamiltonian}. In the main figure we depict the time average of the orders for $2<J t/\hbar<6$ as a function of $U_0$ for the same $F_1$ and $U_1$. The shaded regions indicate the variances of the orders, associated to the dynamics after the quench. In the DMRG simulations of Eq.~\eqref{eq:Hamiltonian} we employ $L=60$ sites, $N=60$ particles, and $\hbar\omega=20\pi J$; the simulations of the effective model~\eqref{eq:effHamiltonian} are performed in an infinite 
scheme~\cite{Schollwock2011}. In both cases we keep up to $M=400$ matrix states.}
\label{fig:2}
\end{figure} 


\emph{Dynamics after switching the double modulation.--} Up to this point we have analyzed the ground-state properties of the 
time-independent model~\eqref{eq:effHamiltonian}. It is however interesting and experimentally relevant to investigate 
the dynamics following the switching of DM. For simplicity we consider at times $t<0$ a large $U_0/J$ at unit filling in absence of any modulation, 
such that the system is in a defect-free MI~(${\cal O}_P^2=1$, ${\cal O}_S^2=0$). At $t=0$ $U_0$ is set to a final value and the sinusoidal DM is abruptly switched on (to values $U_1/\hbar\omega=3$ and $F_1/\hbar\omega=2.4$ in Fig.~\ref{fig:2}), which result in $J_{(0,1)\leftrightarrow (1,0)}, J_{(1,1)\leftrightarrow (0,2)} \ll J_{(1,1)\leftrightarrow (2,0)}$.  We have performed DMRG simulations of the dynamics
employing both the original time-dependent Hamiltonian~\eqref{eq:Hamiltonian} and the effective one~\eqref{eq:effHamiltonian}. As depicted in the inset of 
Fig.~\ref{fig:2}, both models provide identical results, showing the validity of the effective model for describing the dynamics. 
After the quench the creation of defects abruptly reduces ${\cal O}_P^2$ and increases ${\cal O}_S^2$, and then subsequently develop an oscillatory dynamics. Figure~\ref{fig:2} shows the time average of ${\cal O}_P^2$ and ${\cal O}_S^2$ as a function of $U_0$~(the shadowed regions indicate the variance of the orders 
related to the oscillations characterizing the real time dynamics). The average values present a similar qualitative dependence as that expected from the stationary state, 
with growing ${\cal O}_P^2$  and decreasing ${\cal O}_S^2$ for larger $U_0/J$ (see Fig.3 of Ref. \cite{Endres2011} for an analysis of the growth of ${\cal O}_P^2$ for increasing $U_0/J$ without periodic modulations). These results hence show that even an abrupt start of the double modulation transforms an initial defect-free MI into an insulator with non-vanishing time-averaged ${\cal O}_P^2$ and ${\cal O}_S^2$.



\begin{figure}[t]
\centering
\includegraphics[width=\linewidth]{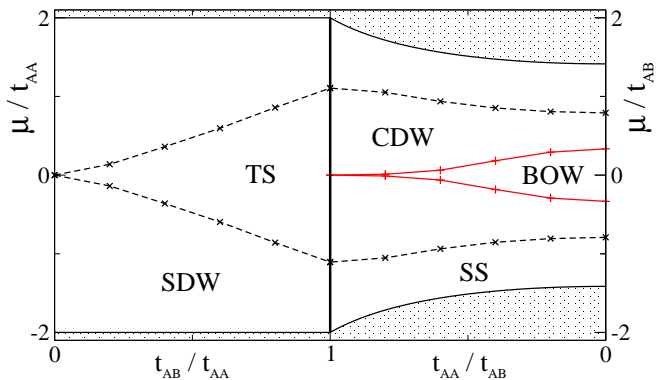}
\caption{(Color online) Phase diagram for the Fermi-Hubbard-model as a function of $\mu/t_{AA}$ and $t_{AB}/t_{AA}$ for $U_0=0$ (see text). The dashed line denotes values of $\mu$ for which the Luttinger-liquid parameter $K_\rho=1$ extracted from the long wavelength behavior of the static charge structure factor \cite{Moreno2011} for $t_{AA}\neq t_{AB}$. This is consistent with the perturbative result from \cite{Aligia1999} for $t_{AA}=t_{AB}$. The size for the BOW-phase corresponds to the charge gap at unit filling. The shaded regions denote the vacuum or the fully occupied state. All results are extrapolated to the thermodynamic limit from open boundary DMRG-calculations with up to $L=144$ sites.}
\label{fig:3}
\end{figure}

\emph{Exploring the complete phase diagram of fermions with correlated hopping.--}
As mentioned above, double modulation allows for  $J_{(1,1)\leftrightarrow (2,0)}/J_{(0,1)\leftrightarrow (1,0)}$ and/or $J_{(1,1)\leftrightarrow (0,2)}/J_{(0,1)\leftrightarrow (1,0)}$ to be larger than $1$, 
and hence for exploring novel quantum phases unreachable with  $U_1=0$ and/or $F_1=0$.   
This is best illustrated by the spin-$1/2$ Fermi-Hubbard model with lattice shaking and modulated  
interactions, $\left[U_0+U_1\cos\left(\omega t\right)\right] \sum_j \hat n_{j\uparrow} \hat n_{j\downarrow}$. Considering for simplicity a spin-independent lattice and 
mirror symmetric hopping,  ${\cal J}_0\left[(F_1+U_1)/\hbar\omega\right] = {\cal J}_0\left[(F_1-U_1)/\hbar\omega\right]$, we reach for sufficiently fast modulations the effective Hamiltonian:
\begin{equation}
H_F\!  =\! U_0\!\sum_j \! \hat n_{j\uparrow}\hat n_{j\downarrow} -\! \!\!\!\!\!\sum_{\langle i,j\rangle,\sigma=\uparrow,\downarrow}\!\!\!\! (\hat c_{i \bar\sigma}^\dag \hat c_{j \bar\sigma}\!  +  {\rm H.c.}) 
{\cal P}(\hat n_{i\sigma},\hat n_{j\sigma}),
\label{eq:HF}
\end{equation}
with $\bar\sigma=-\sigma$, and ${\cal P}(\hat n_{i\sigma},\hat n_{j\sigma})\equiv t_{AA}(1-\hat n_{i\sigma})(1-\hat n_{j\sigma})+t_{BB}\hat n_{i\sigma}\hat n_{j\sigma}
+ t_{AB} \left [ \hat n_{i\sigma}(1-\hat n_{j\sigma})+\hat n_{j\sigma}(1-\hat n_{i\sigma})\right ]$, 
where $\hat c_{j\sigma}$ annihilates a fermion of spin $\sigma$ at site $j$, $\hat n_{j\sigma}=\hat c_{j\sigma}^\dag \hat c_{j\sigma}$,  
$t_{AA}=t_{BB}=J {\cal J}(F_1/\hbar\omega)$, and $t_{AB}=J {\cal J}_0\left[(U_1+F_1)/\hbar\omega\right]$.  Hamiltonian~\eqref{eq:HF} has been extensively studied as 
a model for cuprate superconductors~\cite{Arrachea1994,Arrachea1996,Arrachea1997,Aligia1999,Aligia2000}. Figure~\ref{fig:3} shows 
the grand-canonical phase diagram for $U_0=0$~\cite{footnote-3}. 
For $t_{AB}<t_{AA}$ two spin-massless phases occur~\cite{Aligia1999}: one with dominant triplet superconducting~(TS)~\cite{footnote-TS} correlations 
$\langle {\cal Q}_{0+}^\dag {\cal Q}_{j+}\rangle$, with ${\cal Q}_{j\pm}\equiv c_{j+1\downarrow}  c_{j\uparrow}\pm c_{j+1\uparrow}  c_{\downarrow}$, and another
with dominant spin-density wave~(SDW) correlations, $(-1)^j \langle \hat n_{0-}\hat n_{j -}\rangle$, 
with $\hat n_{j\pm}\equiv  \hat n_{j\uparrow} \pm \hat n_{j\downarrow}$. 
At $t_{AB}=0$ the TS phase has vanishing Drude weight~(Kohn metal~\cite{Arrachea1994,Arrachea1996,Kohn1964}), 
and the SDW at $\bar n\equiv\langle \hat n_{j+}\rangle>1$~($<1$) is a metal without holons~(doublons)~\cite{Arrachea1994}.

Sole modulation of interactions only permits $0\le t_{AB}/t_{AA}\le 1$~\cite{DiLiberto2013}. 
DM allows for  $t_{AB}/t_{AA}>1$, for which three spin-massive phases occur~\cite{Aligia1999}:  
a gapped phase (both in spin and density sectors) at $\bar n=1$  with  bond-ordering wave~(BOW) order, $\langle {\cal  B}_0^\dag {\cal B}_j \rangle$ with 
${\cal B}_j\equiv\sum_\sigma (\hat c_{j+1 \sigma}^\dag \hat c_{j\sigma}+ {\rm H.c.})$, a gapless phase with dominant 
density wave~(CDW) correlations,  $(-1)^j \langle \hat n_{0+} \hat n_{j+} \rangle$, 
and a gapless phase  with dominant singlet-superconducting~(SS) correlations, $\langle {\cal Q}_{0-}^\dag {\cal Q}_{j-}\rangle$.  
Interestingly the gapped BOW phase occurs even for $U_0=0$ due to the effective repulsion induced by the density-dependent hopping.


\emph{Outlook.--} 
DM of cold lattice gases allows for the precise control of selected hopping processes. Such a control permits the realization 
of quantum phases with unconventional properties. In particular, mirror-asymmetric hopping results in insulating 
1D phases with both parity and string orders, which may be revealed using in-situ site resolved imaging~\cite{footnote-4}. 
We have also shown that DM may be used to simulate lattice models in otherwise unreachable 
regimes, as shown for the relevant case of the spin-$1/2$ Fermi-Hubbard model with correlated hopping.

We have considered for simplicity homogeneous systems. With an overall confinement, $V(j)$, that varies slowly enough from site to site, 
 the grand-canonical phase diagram maps into a spatial distribution through the local chemical potential $\mu_j=\mu-V(j)$. 
 In particular, the BOW phase results in a density plateau that may be revealed using single-site resolution~\cite{Sherson2010,Bakr2010}. 
 Moreover, whereas for $U_0=0$ the spin gap opens at $t_{AB,cr}=t_{AA}$ at any $\bar n$, for $U_0>0$, $t_{AB,cr}$ depends on $U_0$ and 
$\bar n$~(for low $U_0/t_{AA}$, $t_{AB,cr}\simeq U_0/(8(\bar n-1)\cos (\pi\bar n/2)-16\sin (\pi \bar n/2)/\pi)$~\cite{Aligia1999}). As a result, 
for fixed $t_{AB}$, $t_{AA}$ and $U_0$, the spin gap opens at a critical $\mu_{cr}$. For an overall confinement there is hence a spatial boundary between 
spin-gapped and spin-gapless phases, which may be revealed by creating spin excitations in the gapless region using a spin-dependent potential, and observing the 
reflection of the excitations at the boundary~\cite{Kollath2005}.

In this article we have focused on the simplest double modulation possible in which both fields were oscillating in phase. A natural and non-trivial extension of this work would be in the analysis of the physics emerging when the two modulations are different and time-reversal symmetry is broken. Another important extension of this work would be in the application of DM to 2D and 3D systems, where the combination of elliptical lattice shaking \cite{Struck2012} and periodically modulated interactions may lead to an even richer physics.

\emph{Acknowledgements.--} We acknowledge support by the cluster of excellence QUEST, the DFG Research Training Group 1729, and the 
SUTD start-up grant (SRG-EPD-2012-045). Part of the computer simulations were carried out on the cluster system of the Leibniz Universit\"at Hannover. Computer simulations for the inset of Fig.\ref{fig:2}, run on the cluster Perseus at the Dept. Th. Ph. of University of Geneva, employed the ALPS libraries~\cite{ALPS,ALPS2} and the code IVAN.

\bibliographystyle{prsty}

\begin{thebibliography}{}
\bibitem{FootnoteHighBands} Higher bands play a role when single-photon or multi-photon processes cannot be neglected. In our model, the typical values of $J/\hbar$ and $U_0/\hbar$ are of the order of $10$ to $100\;{\rm s}^{-1}$ which requires modulations of a $10^3\;{\rm s}^{-1}$. For an interaction modulation with $U_1/\hbar\omega=2$, the oscillation frequency $\hbar\omega< E_{GAP}/10$, where $E_{GAP}$ is gap of the first Bloch band, to ensure that excited bands are not populated. Since $E_{GAP}$ can be at a distance larger than $10^4\;{\rm s}^{-1}$, for our problem the conditions for a single-band description are fulfilled.
\bibitem{footnote-1} We note that, from the choice of $f_U$ and $f_F$, it follows that the Hamiltonian is not invariant under space inversion 
symmetry. In fact for $x\rightarrow -x$ there is no $t_0$ such that the transformation $t\rightarrow t_0\pm t$ leaves the Hamiltonian unchanged. 
The Hamiltonian is however still invariant for time inversion symmetry $t\rightarrow -t$.
\bibitem{footnote-2} One may cancel the other relevant processes at low filling by properly choosing 
$F_1$ and $U_1$.
\end{thebibliography}

\end{document}